# Hardware Implementation of Adaptive Watermarking Based on Local Spatial Disorder Analysis

Mohsen Hajabdolahi[a], Nader Karimi[a], Shahram Shirani[b], Shadrokh Samavi[a,b],
[a]Department of Electrical and Computer Engineering, Isfahan University of Technology, Isfahan, Iran.
[b]Department of Electrical and Computer Engineering, McMaster University, Hamilton, Canada

**Abstract**
With the increasing use of the internet and the ease of exchange of multimedia content, the protection of ownership rights has become a significant concern. Watermarking is an efficient means for this purpose. In many applications, real-time watermarking is required, which demands hardware implementation of low complexity and robust algorithm. In this paper, an adaptive watermarking is presented, which uses embedding in different bit-planes to achieve transparency and robustness. Local disorder of pixels is analyzed to control the strength of the watermark. A new low complexity method for disorder analysis is proposed, and its hardware implantation is presented. An embedding method is proposed, which causes lower degradation in the watermarked image. Also, the performance of proposed watermarking architecture is improved by a pipe-line structure and is tested on an FPGA device. Results show that the algorithm produces transparent and robust watermarked images. The synthesis report from FPGA implementation illustrates a low complexity hardware structure.
**Keywords**: Information hiding; watermarking; low complexity; hardware implementation; FPGA

## 1. Introduction

With the ever-increasing of communication of digital data, the security of digital content has been an essential problem. Digital watermarking, as a branch of information security, has been extensively studied for digital media, such as images, voice, and video files. Digital watermarking embeds information about the intellectual property of a digital media inside the media itself [1, 2].

The implementation of watermarking could be performed by different means, such as software, hardware, embedded controllers, DSP processors, etc. [3]. In some applications of digital watermarking, real-time performance is required. Hardware implementation can be considered as an approach to meet the real-time demands [4]. Also, this approach can be used as an accelerating technique for the implementation of watermarking algorithms. Application-specific integrated circuits (ASIC) and field-programmable gate arrays (FPGA) implementations are two efficient ways for hardware implementation where each one has its advantages and disadvantages. From another perspective, watermarking can be performed on spatial and transform domains. Direct pixel modification is an example of watermarking in the spatial domain. Changes of coefficients of discrete cosine transform (DCT) or Wavelet are examples of watermarking in transform domains.

There are many studies in the field of hardware design and implementation of digital image watermarking algorithms [5]. In practical applications, there are different constraints for hardware design, such as high reliability, low cost, low complexity, and ease of implementation with the existing consumer electronic devices. Due to these constraints, usually watermarking methods that are chosen for the hardware implementation are spatial domain methods.

In [6], a circuit is designed which could embed data in images for robust and fragile watermarking. A ternary watermark system is used for invisible embedding, which adds scaled gray values corresponding to the intensities of pixels of a neighborhood. In [7], a very large scale integration (VLSI) layout is presented, which is capable of invisible as well as fragile watermarking in the spatial domain. They use a linear feedback shift register (LFSR) for the watermark generation, and structural details of the circuit are presented. In [6, 7], it is mentioned that a disadvantage of hardware watermarking algorithms is that the processes have to be done in a pixel by pixel manner.

In [8], a VLSI design for watermarking is developed, which uses the spatial domain for embedding purposes. In that paper, a simple LSB replacement scheme is offered, which operates on 8-bit pixels, and the design is done at the layout level. Spatial domain watermarking offers low computational overhead as compared to that of the frequency domain. It is claimed in [8] that their LSB replacement scheme is robust against lossy image compression processes and some filtering alterations too. In [9] it is attempted to develop a VLSI architecture for a high-performance watermarking chip. They try to perform transparent color image watermarking in the special domain by using a genetic algorithm (GA). The genetic algorithm is used for finding a search space and optimizing the intensity to best fit for image watermarking. This implementation of GA, for image adaptivity, is done at the cost of high hardware use and complexity. In [10, 11] an FPGA and ASIC designs of watermarking in the spatial domain are presented. With the definition of a connectivity preserving criterion, watermark embedding is performed on a pixel. For embedding, LSB of a pixel is flipped such that the flipping would not destroy the connectivity between the pixels and create extra clusters as well. In [12], an FPGA based implementation of an invisible- spatial domain watermarking encoder is presented .For watermark embedding, the relation of the original pixel and its neighboring pixels is considered, and according to the watermark bit, the pixel values are changed. A finite state machine (FSM) is used for hardware implementation of the control unit, and the algorithm is implemented on an FPGA device.

Among the hardware watermarking techniques in the spatial domain, some previous studies are based on transform domains. Potentially conventional transform-based watermarking methods have high hardware complexity. Hence to have a design with acceptable hardware specifications, the hardware optimization techniques must be taken into account. In [13], hardware architecture is proposed to transform domain watermarking using a quantization approach. Their design uses pipelining, and it is attempted to optimize the performance of the pipeline. The main objective of the study of [13] is to propose VLSI architecture for a blind image watermarking chip. In [14], authors intend to combine the advantages of both spatial and transform domains for a relatively low computational complex implementation. This approach is implemented on FPGA. In [15], both watermarking and encryption techniques are implemented in the DCT domain on FPGA. In [16], a wavelet domain watermarking method is presented for FPGA implementation. They attempt to reduce the high hardware complexity of wavelet implementation by a special insertion technique. In [17], DCT based visible and invisible watermarking methods are presented. Hardware optimization techniques such as a five-stage pipeline to improve performance and low power techniques such as dual voltages, dual-frequency, and clock gating are used to reduce power consumption. In [18], an FPGA implementation of DCT based watermarking is presented. By reducing the number of multipliers and adders in DCT calculation, a five-stage pipeline is implemented.

The advent of FPGAs provided an easy and quick prototyping environment. These devices can be programmed for different functionalities without all difficulties that exist in custom integrated circuit design and manufacturing [2]. Most hardware implementations of transform-domain watermarking methods have high complexities. On the other hand, hardware implementations of spatial-domain watermarking methods are not usually adaptive and do not consider image content for embedding strengths.

In this paper, an adaptive watermarking method and its enhanced version in the spatial domain are proposed, which are compatible with efficient hardware implementation. A simple criterion for local pixel disorder analysis is introduced and compared with three complex criteria. Simulation results show the correct identification of crowded regions with pixel disorders. Based on this disorder criterion, embedding is performed with different strength levels. Also, the implementation of this algorithm on FPGA is performed that produce good quality watermark images. Adaptive properties of the algorithm cause watermarking to cause changes that have minimal effects on the human visual system. Our main contributions are a) a simple and low complexity disorder analysis for detection of crowded regions, b) an adaptive watermarking based on the local disorder of pixels, c) a modified watermarking to improve the signal to noise ratio and visual quality of the watermarked image, d) efficient pipe-line implementation of adaptive watermarking.

The rest of the paper is organized as follows: Section 2 describes the details of the proposed adaptive watermarking algorithm. The proposed pipelined watermark embedding circuit is explained in Section 3. Experimental results are presented in Section 4, and Section 5 is dedicated to the conclusion of the paper.

## 2. Proposed Adaptive Watermarking

We can consider natural images as having two types of regions. The first category is for regions that are relatively smooth or "ordered." The second category belongs to regions that are congested or "disordered." In general, spatial domain watermarking is the process of changing pixel values such that minimal effects are caused in the visual quality of the image. The sensitivity of the human visual system (HVS) to changes that are caused in a region of the image depends on how congested that region is. In adaptive watermarking methods, more extensive changes could be performed in pixel values, and still, HVS could not realize them. Hence, an adaptive method should first distinguish between ordered and disordered regions of the image. Then watermark data can be embedded with high strength in disordered areas while for embedding watermark data in ordered regions, smaller changes should be applied.

### 2.1. Embedding method

In a watermarking process, the visual quality of the output plays an essential characteristic of the procedure. Hence, our watermarking algorithm is implemented such that to cause minor effects on the visual quality of the image. The Block diagram of the proposed watermarking system is illustrated in Fig. 1.

The algorithm, as shown in Fig. 1 has two stages of Congestion analysis and Adaptive embedding. Initially, the algorithm performs a congestion analysis to distinguish between ordered and disordered regions. Then depending on whether the image block under investigation is ordered or disordered, an appropriate embedding is performed by the Adaptive embedding stage. In the following, each of the two stages of Congestion analysis and Adaptive embedding is explained.

### 2.2. Congestion analysis

In this paper, a region is considered as congested and disorder if variations in that region are relatively high. In a congested region of an image, the human visual system (HVS) is not capable of detecting artifacts that are created by the watermarking process. On the other hand, HVS is sensitive to changes that are caused by less congested and smooth areas. Different methods can analyze the amount of disordering in an image. Some of these methods of congestion analysis methods are listed briefly in the following.

**Edge detection:** The presence of edge pixels in a block could be an indication of the disorder among pixels of that block [19]. Such regions are proper for robust embedding without causing psycho visual effects [20].

**Entropy analysis:** Entropy can be used to measure "disorder" [21, 22]. Entropy can be used as a criterion indicating information contained in an image [23].

**Transform space:** Coefficients of a frequency domain transform method, such as DCT, could also be used to find out whether a block of an image is smooth or disordered. Coefficients of DCT of a block of an image indicate the presence of different frequencies and pixel disorder in that block [24-27].

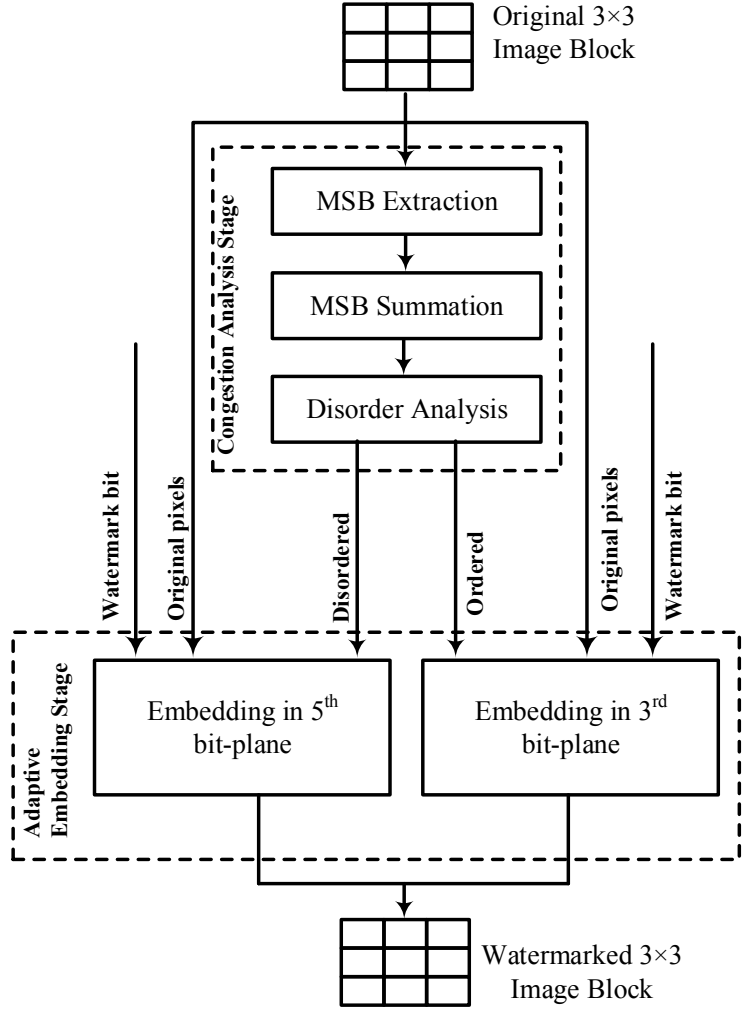

Fig. 1. General system structure of the proposed adaptive watermarking method.

**Proposed disorder analysis:** All of these existing congestion analysis methods that we mentioned have computational complexities that are not suited for hardware implementation. Therefore, we propose a very simple congestion analysis method that uses the most significant bit-plane (MSB) of the pixels of a block to decide whether the block is ordered or disordered. For this purpose, the image is partitioned into 3×3 non-overlap blocks. An example of a block is shown in Fig. 1. How "zeros" and "ones" are distributed among MSBs of pixels in a block, could indicate the amount of disorder or congestion. This is done, as shown in Fig. 1, by two boxes called MSB Extraction and MSB Summation. Let us assume that $P$ is the block under analysis, and the sum of the MSBs in the 3×3 block is $S$. Based on the value of $S$ block $P$ is classified as one of the two types:

1) If $S \in \{0, 1, 2, 3, 7, 8, 9\}$, then block $P$ is considered to be in a smooth region, and this block is of type "ordered."
2) If $S \in \{4, 5, 6\}$, then $P$ is in a congested region, and the block is of type "disordered."

This analysis is very hardware friendly, and it produces acceptable accuracy as compared to the other mentioned methods. We will show that the experimental results verify this claim.

## 2.3. Adaptive embedding

A block of pixels of the image is considered. All of the bits in the MSB bit-plane of this block are added together. The sum is then compared with the threshold value. A block is labeled as ordered or disordered type. In ordered blocks, which are supposed to be smooth, embedding is done in the lower significant bit-planes (shown in Fig. 1 as "embedding in the 3$^{rd}$ bit-plane"). This is done to cause a minimal change in a smooth region. Also, in the disordered areas that are congested, watermarking is done in higher bit-planes (shown in Fig. 1 as "embedding in 5$^{th}$ bit-plane"). This is a stronger embedding which results in more robust watermarking. If we number the bit-planes from 1(least significant) to 8 (MSB), then in smooth and congested regions, embedding is done in the 3$^{rd}$ and 5$^{th}$ bit-planes, respectively. Hence based on the content of the image, the embedding strength is changed. We call this algorithm as the "basic adaptive" method.

## 2.4. Enhanced adaptive embedding

In the previous section, the basic adaptive embedding was explained which 3$^{rd}$ and 5$^{th}$ bit-plane were selected for smooth and crowded areas, respectively. The embedding process changes the values of pixels and influences the visual quality of the image. It is possible to reduce the severity of changes by modifying the embedding process. For this purpose, all 8-bit positions of the pixels of a 3×3 block are illustrated in Fig. 2. The initial embedding bit-planes for smooth and congested blocks are shown. Also, the bit-plane, next to the embedding bit-plane, is used for enhancement purposes. In the enhancement process, in addition to the main adaptive embedding, the inverse of the embedded watermark bit is embedded in the next lower bit-plane. In Fig. 2, the original embedding bit-plane is shown in gray, and the enhancement position is shown in green. For example, if we are embedding 1 in the 5$^{th}$ bit-plane of a block, then we embed 0 in its 4$^{th}$ bit-plane. We will prove that the visual quality, in terms of peak signal to noise ratio (PSNR) and the structural similarity (SSIM) index, can be improved.

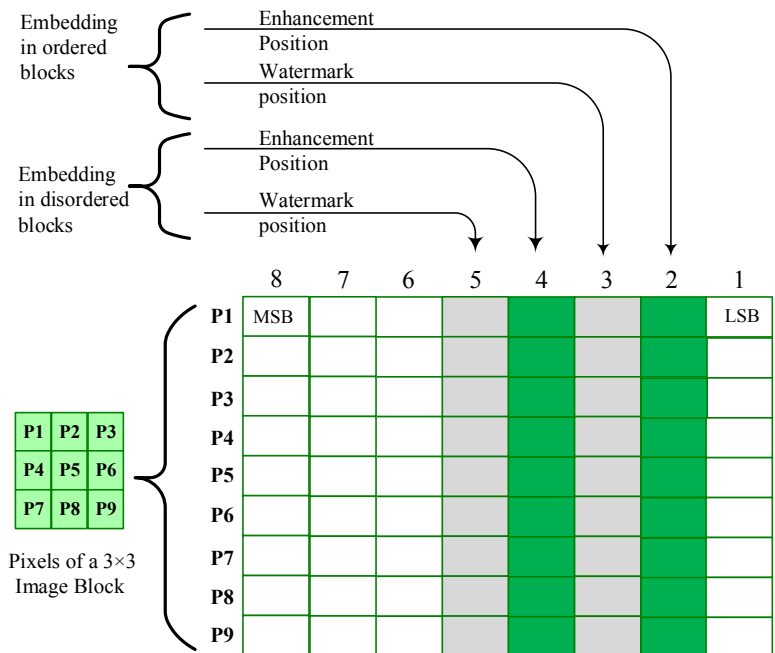

Fig. 2. Enhancement of the adaptive watermarking. Watermark bit positions are shown in gray and enhancement bit positions are shown in green.

We propose that if in the embedding process a bit $w$ is embedded in the bit-plane $(i + 1)$ then by inserting $\bar{w}$ in the $(i)$th bit-plane, we would observe improvements in PSNR values. To indicate the effect of the enhanced version of the algorithm, let us look at the definition of PSNR as:

$$\text{PSNR} = 10 \log \left( \frac{(maximum\ pixel\ intensity)^2}{MSE} \right)$$

where the mean square error (MSE) is calculated using the difference between every two corresponding pixels in the original and watermarked versions of an $M \times N$ image as:

$$\text{MSE} = \frac{1}{MN} \sum_{n=1}^{N} \sum_{m=1}^{M} (D(m,n))^2$$

$$D(m,n) = I(m,n) - \tilde{I}(m,n)$$

where $I(m,n)$ and $\tilde{I}(m,n)$ are two corresponding original and watermarked pixels at coordinates $(m,n)$. Suppose we chose a pixel with random intensity, and then a random watermark bit $w$ is embedded in the bit position $(i + 1)$ of this pixel. In our basic embedding method bit $b_{i+1}$ of the pixel is replaced by $w$. Then, depending on the values of $b_{i+1}$ and $w$, there is a probability $P$ that due to this basic embedding method the pixel is changed by an amount of $D_{basic}^{(k)}$. Superscript $(k)$ indicates different situations that $b_{i+1}$ and $w$ could have, which determines the magnitude of change. We also use the notation $D_{enh}^{(k)}$ to refer to possible changes that could be formed in a pixel due to the enhanced embedding. We know that MSE is based on the square of changes that are occurred in pixels. Hence, for the basic algorithm, MSE is closely related to the expected value of $(D_{basic})^2$, namely $E[(D_{basic})^2]$. We want to show that the enhanced version of the algorithm has a lower expected value of the square of changes than its basic version, which means $E[(D_{basic})^2] > E[(D_{enh})^2]$. If this is true, then the PSNR values of the enhanced version would be higher than those of our basic adaptive method.

We start by calculating $E[(D_{basic})^2]$ of our basic algorithm. The absolute value of difference that is caused in a pixel is $\left| D_{basic}^{(k)} \right|$ and is calculated as:

$$\left| D_{basic}^{(k)} \right| = \begin{cases} \left| D_{basic}^{(1)} \right| = 0 & if\ b_{i+1} = w \\ \left| D_{basic}^{(2)} \right| = 2^{i+1} & if\ b_{i+1} \neq w \end{cases} \quad (1)$$

$$k = \{1,2\}, \quad 0 \leq i \leq 6$$

For random values of $w$ and $b_{i+1}$ the probability of occurrence of each of the above two cases in Equation (1) is equal to 0.5. Hence, the expected value of $(D_{basic})^2$ is calculated as:

$$E[(D_{basic})^2] = \sum_{k=1}^{2} P(D_{basic}^{(k)}) \times (D_{basic}^{(k)})^2 = 2 \times 2^{2i}$$

Now we calculate the expected value of changes that could occur in a pixel by our enhanced embedding method, which is called as $E[(D_{enh})^2]$. In this enhanced version of the algorithm, a watermark bit $w$ is embedded into two bits of a pixel, $b_{i+1}b_i$, by replacing these two bits by $w\bar{w}$. When a random $w$ is embedded in a pixel using the enhance method, the absolute value of change that occurs in the pixel is called $\left| D_{enh}^{(k)} \right|$ which is shown by the following 4 cases:

$$\left| D_{enh}^{(k)} \right| = \begin{cases} \left| D_{enh}^{(1)} \right| = 2^i & if\ (b_{i+1} = b_i) and (b_{i+1} = w) \\ \left| D_{enh}^{(2)} \right| = 0 & if\ (b_{i+1} \neq b_i) and (b_{i+1} = w) \\ \left| D_{enh}^{(3)} \right| = (2^{i+1} - 2^i) & if\ (b_{i+1} \neq b_i) and (b_{i+1} \neq w) \\ \left| D_{enh}^{(4)} \right| = 2^{i+1} & if\ (b_{i+1} = b_i) and (b_{i+1} \neq w) \end{cases} \quad (2)$$

$k = \{1,2,3,4\}, \quad 0 \leq i \leq 6$

For random values of $w$, $b_{i+1}$, and $b_i$ all of the 4 cases of Equation (2) have an equal probability of occurrence of ¼. Therefore, the expected value of $(D_{enh})^2$ for the enhanced embedding method would be:

$$E[(D_{enh})^2] = \sum_{k=1}^{4} P(D_{enh}^{(k)}) \times (D_{enh}^{(k)})^2 = (\frac{3}{2}) 2^{2i}$$

This shows that $E[(D_{basic})^2] > E[(D_{enh})^2]$. Therefore, we expect to see higher PSNR values using the enhanced version of the embedding method.

### 2.5. Extraction method

The message, which exists in a watermarked image, is extracted in two steps as described in the following.

**Step1:** The image is partitioned into non-overlapped 3×3 blocks. For each block, congestion is determined by summing the MSB bit-plane.

**Step2:** In smooth blocks 3rd bit-plane contains the watermark bit, and in congested blocks, the embedded bit is in the 5th bit-plane. A voting process on the number of 1s in the intended bit-plane would reveal the extracted bit. If the number of 1s is greater than 4, then the extracted bit is 1; otherwise, it is 0.

## 3. Proposed Pipelined Embedding

In this section, we offer the formation of a pipeline structure for watermark embedding of an image. This task is achieved by taking an image into a pipeline row by row. Then after three rows of the image are read, watermark embedding is performed. Subsequently, another three rows are read and so forth. The architecture of the proposed pipeline is shown in Fig. 3.

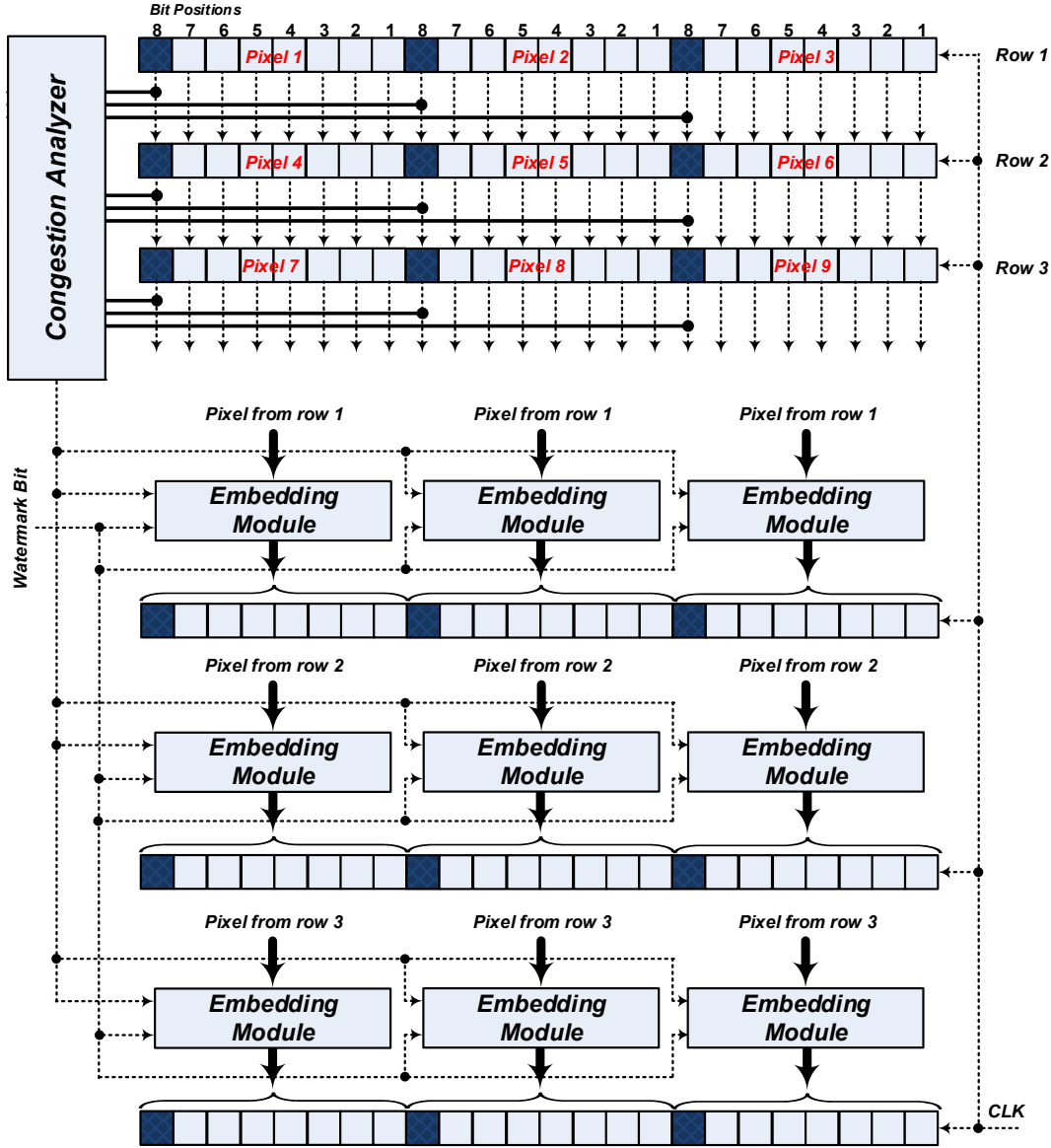
Fig. 3. Proposed pipeline design for congestion analysis and watermark embedding

The two main hardware modules of Fig. 3 are the congestion analyzer and the embedding module. Efficient hardware implementation for each part is considered and discussed in the following sub-section.

### 3.1. Congestion Analyzer

This module calculates the amount of congestion and categorizes a block into two types of smooth or congested. A hardware implementation of the congestion analyzer is proposed in Fig. 4. The two essential elements of this module are a 9-bit adder and a type indicator circuit which are explained in the followings:

**9-bit adder architecture:** In the proposed watermarking algorithm, the adder is an important part of the design. All elements in the MSB bit-plane of this mask must be efficiently added. This task is implemented by a 9-4 compressor is used, which adds all of the 9 MSB bits of a 3×3 block to form a 4-bit number [28]. As shown in Fig. 4, in this adder architecture, only 5 full adders (FA) and 2 half adders (HA) are used, which make it proper for real-time implementation.

**Type indicator:** After adding all of the bits in the MSB bit-plane, the congestion of the block is categorized by a type indicator. This indicator circuit should decide whether the result of the adder is a member of the set {0, 1, 2, 3, 7, 8, 9}, which indicates an ordered block, or whether the addition result is a member of the set {4, 5, 6}, which makes the block as disordered type. For this purpose, we implemented the needed type indicator based on the structure shown in Fig. 4. A logic 1 at the output of this circuit indicates disordered type blocks. This circuit only uses three logic gates and is suitable for hardware implementation. Conversely, a logic 0 at the output of this circuit would indicate an ordered block.

### 3.2. Embedding Module

Each of the embedding modules that are shown in Fig. 3 receives a signal from the congestion analyzer and embeds a watermark bit in an appropriate bit position of a pixel. For hardware implementation of the embedding module, two multiplexers are used to select between two embedding positions ($3^{rd}$ or $5^{th}$ bit positions). The hardware architecture for the embedding process is shown in Fig. 5. For simplicity in representation, in Fig. 5 the architecture for one pixel is shown. The selector lines for the multiplexers are activated by the output of the type indicator. Based on the type of block, one of the multiplexers allow the watermark bit to be embedded in the pixel and the other multiplexer simply passes the original bit to its output. This mechanism chooses between $3^{rd}$ and $5^{th}$ bit positions.

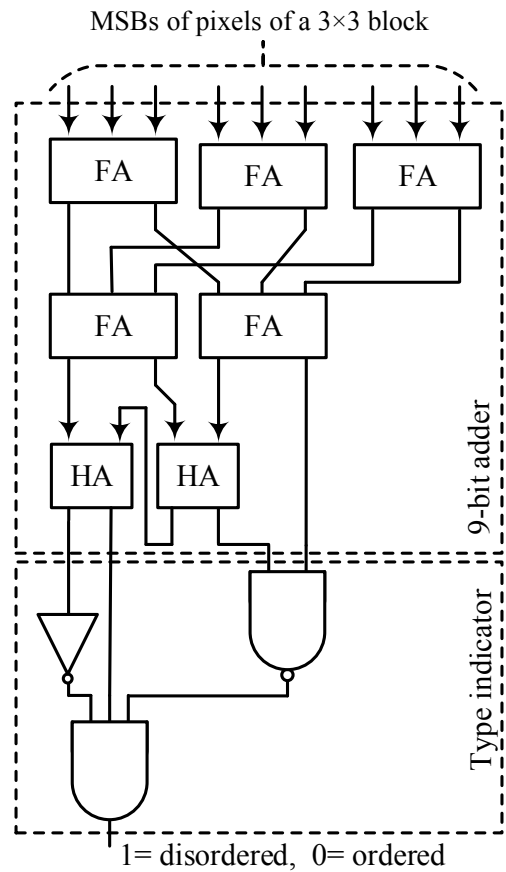

Figure. 4. Hardware implementation of the congestion analyzer consisting of a 9 to 4 compressor [28] and a type indicator circuit.

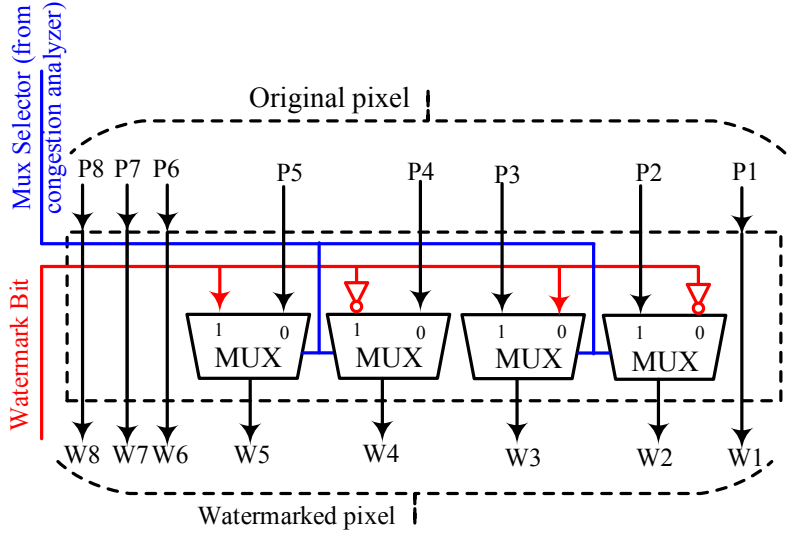

Fig. 5. Logic structure of the embedding module for the enhanced adaptive method for embedding in 3rd or 5th bit-planes.

## 4. Experimental Results

The correctness and the embedding performance of the proposed watermarking algorithm are verified by MATLAB implementation. A binary sequence of M bits is embedded in an image that has M non-overlapped 3×3 blocks. Also, simulations for hardware implementation are performed by the XILINX ISE design suite, and its results are compared with the MATLAB results. We first explain some of the results that are obtained from software simulations and then present the results from the hardware implementation of the algorithm.

### 4.1. Software Simulation

**Congestion Analysis:** Four congestion analysis methods based on entropy, edge, and DCT transform, and the proposed congestion analysis, were explained in Section 2.2. These methods are all simulated find strength and weaknesses of the proposed congestion analysis method. For each image, we perform our proposed congestion analysis, which is based on counting the number of 1s in the MSB bit-plane of each block. Then we count the number of disordered blocks and call this number as $N_D$. We use the other three congestion analysis methods and based on each of the criteria we sort all of the blocks and chose the top $N_D$ blocks. To implement DCT congestion analysis in pixels of a block, we compute DCT of each non-overlapped block. Then high-frequency coefficients are set to zero, and inverse transform is performed. Mean square error (MSE) between the original block and the reconstructed version is computed. All of the blocks are sorted based on their MSE value, in descending order. The top $N_D$ blocks are chosen as disordered blocks. Hence, the number of congested blocks is the same as our proposed congested analysis method, but the locations of such blocks may differ.

Similarly, we can use entropy for congestion analysis. The entropy of each non-overlapped block is calculated, and the entropy value is assigned to that block. We sort all of the blocks, in descending order, based on their entropy values and the top $N_D$ blocks are considered. For congestion analysis using edge detection, the Canny edge detector is applied to the image. Then in each block, the number of edge pixels is counted and is considered as the congestion amount for that block [20]. Then all of the blocks are sorted in descending order and the top $N_D$ blocks are chosen as disordered blocks.

Using these four congestion analysis methods, the proposed adaptive embedding process, as discussed in Section 2.3, is performed for each of the congestion analysis methods. We performed four experiments, and in all of them, everything was similar except for the congestion analysis method. Each experiment consisted of basic adaptive embedding and a JPEG attack with a quality of 80.

Table 1 shows the results of four watermarking experiments with different congestion analysis methods. Then to show the level of robustness of each method, we compare the extracted watermark with the original watermark by calculating the normalized correlation (NC) between them. Also, to show the quality of the embedded image, we use PSNR and SSIM. These experiments are performed on six sample pictures, and results are illustrated in Table 1 in terms of robustness and watermarked image quality. Table 1 shows that watermarking with any of the four congestion analysis methods would lead to relatively similar results in terms of robustness and transparency. It is useful to note that the congestion analysis by counting 1s in the MSB bit-plane has a significantly lower hardware complexity in comparison with DCT calculation, entropy analysis, or canny edge detection.

**Image Watermarking Process:** Another set of software simulations are performed to test the functionality and performance of the proposed method. This is done by partitioning the image into non-overlapped 3×3 blocks, and a random binary sequence, equal to the number of blocks in the image, is considered as the watermark message. Four different sets of experiments are performed, and the results are shown in Table 2. In the first set of experiments, the message is embedded in the $3^{rd}$ bit-plane of the blocks. The quality of the watermarked image is computed based on its PSNR and SSIM values. Then JPEG attacks with quality factors of 90 and 80 are implemented on the watermarked image. Bits of some lower bit-planes of the converted image change due to the JPEG conversion process. The embedded message is then extracted from the JPEG image. Normalized correlation (NC) of the extracted message with the original binary sequence is calculated as a measure of the robustness of the method. The other three sets of experiments are only different from the first one in their embedding method. The second set of experiments embeds the message in the $5^{th}$ bit-plane of all blocks. The third set of experiments uses our proposed basic adaptive process, which embeds in the $3^{rd}$ or $5^{th}$ bit-plane depending on the congestion of each block. The last sets of experiments that are reported in Table 2 use the enhanced embedding method. Overall we see from Table 2 that embedding the message only in the $3^{rd}$ bit-plane would result in a high quality watermarked image with PSNR values in the range of 42dB. But this high quality is at the expense of low robustness. We see NC values as low as 0.59.

On the other hand, embedding only in the $5^{th}$ bit-plane would result in low-quality images, but the robustness is enhanced. We see PSNR values are about 36dB, but the lowest NC value is about 0.95. Experimental results of the proposed methods combine the advantages of the two bit-planes and come up with a tradeoff between image quality and robustness.

Results of Table 2 shows that the proposed basic adaptive embedding produces better PSNR values than the results of embedding in the $5^{th}$ bit-plane, and its NC values are better than embedding in the $3^{rd}$ bit-plane. We see 3 to 4dB increase in quality of the image, and NC values are increased by at least 0.3. Furthermore, the enhanced version of the proposed adaptive watermarking has improved both the quality and robustness of the images.

Also, in Fig. 6, extracted logo in the presence of three attacks is illustrated. Visual quality (PSNR) and normalized correlation (NC) from Fig. 6 show the robustness of the proposed algorithm for different attacks. Robustness for the cases of salt and pepper, as well as median attacks, is more apparent.

Table 1. Comparison of watermarking using proposed congestion analysis and three other methods.

| Image | | Metric | DCT | Entropy | Edge | Proposed |
|---|---|---|---|---|---|---|
| Lena | 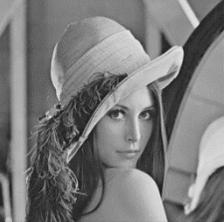 | PSNR (dB) | 39.95 | 38.82 | 39.53 | 40.66 |
| | | SSIM | 0.9640 | 0.9493 | 0.9478 | 0.9573 |
| | | NC | 0.9017 | 0.8333 | 0.8311 | 0.8866 |
| Cameraman | 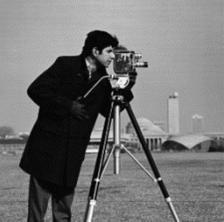 | PSNR (dB) | 40.92 | 39.92 | 41.00 | 41.06 |
| | | SSIM | 0.9627 | 0.9537 | 0.9549 | 0.9544 |
| | | NC | 0.9219 | 0.8731 | 0.8859 | 0.9092 |
| Barbara | 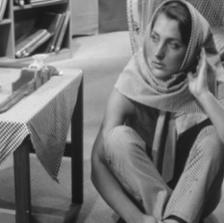 | PSNR (dB) | 37.48 | 37.59 | 38.45 | 39.95 |
| | | SSIM | 0.9533 | 0.9469 | 0.9467 | 0.9685 |
| | | NC | 0.8746 | 0.7808 | 0.7958 | 0.8198 |
| Jet Plane | 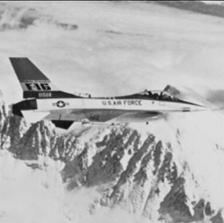 | PSNR (dB) | 43.75 | 40.24 | 40.13 | 42.13 |
| | | SSIM | 0.9592 | 0.9424 | 0.9284 | 0.9545 |
| | | NC | 0.7686 | 0.7810 | 0.7717 | 0.7887 |
| House | 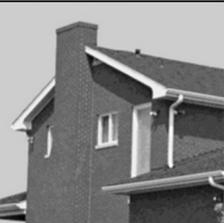 | PSNR (dB) | 45.94 | 41.97 | 41.62 | 43.87 |
| | | SSIM | 0.9703 | 0.9547 | 0.9073 | 0.9663 |
| | | NC | 0.8835 | 0.8553 | 0.7920 | 0.8910 |

### 4.2. Hardware Implementation

For hardware evaluation of the proposed embedding logic, a Xilinx Spartan 3 FPGA device is chosen. The proposed adaptive embedding algorithm is described by VHDL and then it is synthesized on FPGA. For implementing the whole watermarking process on an image, an xc4vlx200 FPGA device, which is a member of the Xilinx virtex4 family, is used. All of the required embedding processes are designed in structural format. Three internal RAMs, for the input image, output image and the watermark message, are used. In this implementation, to achieve high processing throughput, the whole process is pipelined. The process is synchronized so that with every clock pulse, one image line is read from the RAM. Then, after three clock pulses, the congestion analysis and the subsequent embedding process are performed. For an $N \times N$ image, after the initialization phase, in each clock pulse, one row of the image is read. The pipeline is filled in six clock pulses. Afterward, the information embedding is performed in each clock pulse by the appropriate watermarking modules. Therefore $N + 6$ clock pulses are needed to finish the adaptive watermarking procedure. Results are prepared after six clock pulses, and the watermark image is written into its dedicated internal RAM.

Table 2. Robustness and transparency results for JPEG attacks.

| Watermark Method | Watermarking Quality Metric | Cameraman | Lena | House | Barbara | Jet plane |
|---|---|---|---|---|---|---|
| Embedding only in bit-plane 3 | SSIM | 0.9672 | 0.9743 | 0.9656 | 0.9822 | 0.9607 |
| | PSNR (dB) | 44.83 | 45.30 | 44.50 | 45.17 | 45.21 |
| | NC for JPEG attack (Q=90) | 0.8821 | 0.8221 | 0.8814 | 0.8063 | 0.7782 |
| | NC for JPEG attack (Q=80) | 0.7853 | 0.7132 | 0.8049 | 0.7050 | 0.6770 |
| Embedding only in bit-plane 5 | SSIM | 0.6520 | 0.7018 | 0.5521 | 0.7660 | 0.6078 |
| | PSNR(dB) | 31.07 | 30.35 | 33.50 | 30.33 | 30.03 |
| | NC for JPEG attack (Q=90) | 1 | 1 | 0.9996 | 1 | 1 |
| | NC for JPEG attack (Q=80) | 1 | 0.9977 | 0.9972 | 0.9925 | 0.9940 |
| Basic Adaptive | SSIM | 0.9308 | 0.9429 | 0.9338 | 0.9566 | 0.9330 |
| | PSNR (dB) | 37.21 | 38.06 | 39.74 | 37.27 | 39.35 |
| | NC for JPEG attack (Q=90) | 0.9677 | 0.9557 | 0.9657 | 0.9287 | 0.9192 |
| | NC for JPEG attack (Q=80) | 0.9092 | 0.8559 | 0.9014 | 0.8326 | 0.7773 |
| Enhanced Adaptive | SSIM | 0.9544 | 0.9573 | 0.9663 | 0.9685 | 0.9545 |
| | PSNR (dB) | 41.06 | 40.66 | 43.87 | 39.95 | 42.13 |
| | NC for JPEG attack (Q=90) | 0.9625 | 0.9595 | 0.9671 | 0.9272 | 0.9237 |
| | NC for JPEG attack (Q=80) | 0.9092 | 0.8866 | 0.8910 | 0.8198 | 0.7887 |

In Table 3, the comparisons between the proposed watermarking and related works are illustrated. These simulation results show that the proposed watermarking methods have real-time properties both into integrating with consumer electronic devices.

## 5. Conclusion

Hardware architecture of an adaptive image watermarking system, for image authentication, was presented in this paper, for real-time applications. First, a simple and efficient adaptive image watermarking method, suitable for hardware implementation, was proposed. Next, an enhancement design was introduced to create an adaptive method. Then an efficient implementation of the proposed adaptive embedding algorithm was designed. Finally, a pipeline method was used to speed up the watermarking algorithm. Design specifications from FPGA implementation showed that the algorithm could be used in real-world applications with satisfactory performance. Only a low percentage of the total area of the FPGA was used for this hardware design. Although the implementation of watermarking algorithms in the special domain usually has low hardware complexity, but the robustness of such implementations is a significant deficiency. Our enhanced adaptive watermarking tried to solve this problem by detection of regions in which watermark information was preserved against attacks. An error compensation method was employed to increase the visual quality and robustness of the results. For example, we achieved close to 40dB PSNR values for a typical embedded image. Also, the robustness against the JPEG attack was in an acceptable NC range of 0.7 to 1.

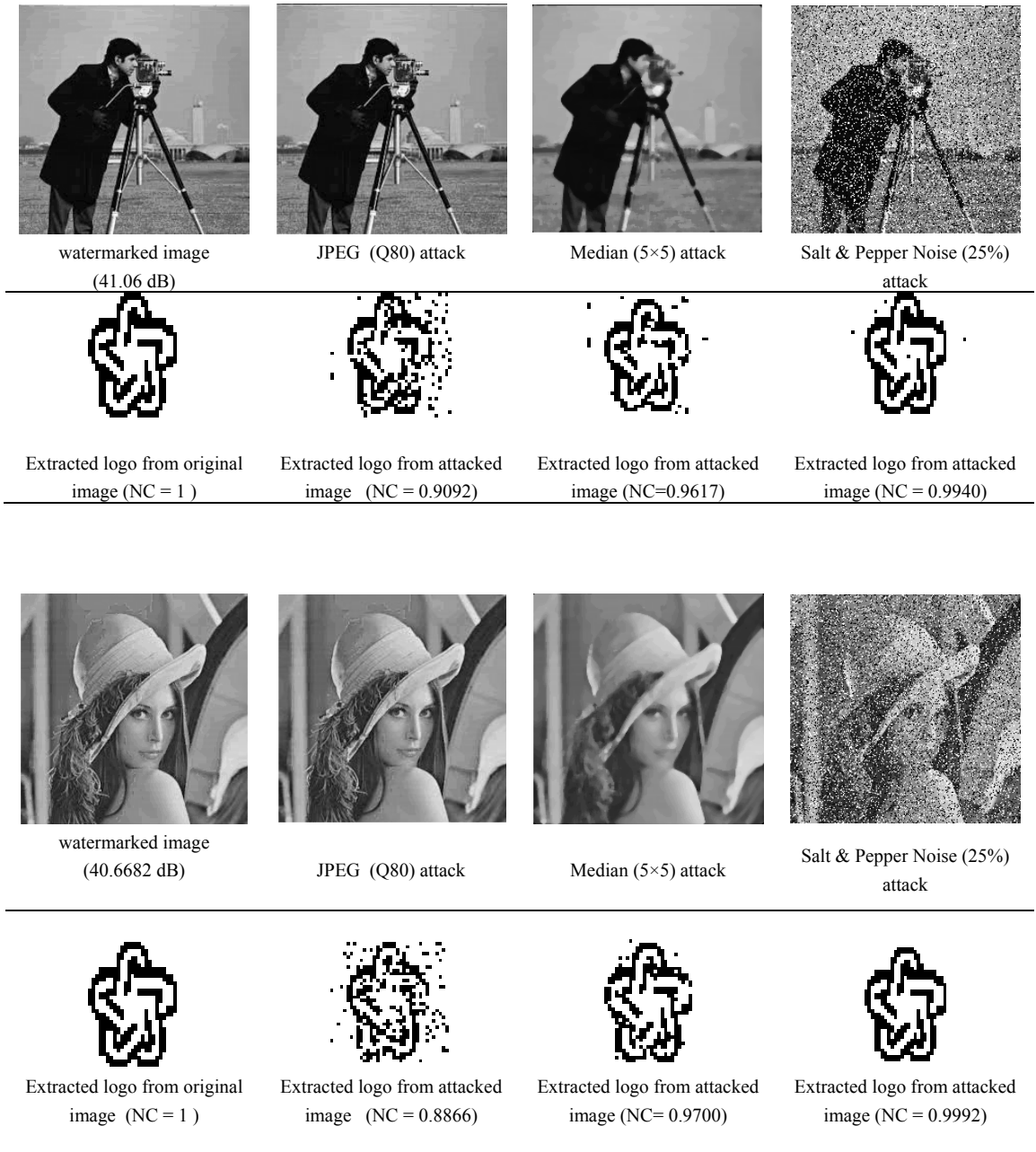

Fig. 6. Extraction of logo under different attacks.

Table 3. Comparison between hardware implementation of the proposed method with other compatible methods.

| Method | Type | Domain | Target Device | Area | Delay or Frequency | Capacity |
|---|---|---|---|---|---|---|
| Ref [10] | Image, Invisible | Spatial | Xilinx VIRTEX-E | 279 FF 788 LUT 457 slices | Not Reported | 1/9 Bit per Pixel |
| | Image, Invisible | Spatial | ASIC 0.35 μm | 1325.026 $\mu m^2$ | 100 MHz | 1/9 Bit per Pixel |
| Ref [16] | Video, Invisible | Wavelet | Altera Stratix II | 22,152 ALUT 5368 Register 72 block DSP Not reported | 8.2 ms / frame | One image in video stream |
| Ref [12] | Image, Invisible | Spatial | Xilinx VIRTEX-II | 959 slices | 12 ns | 1/9 Bit per Pixel |
| Ref [18] | Image, Invisible/Visible | DCT | Xilinx Spartan 3 | 6734 LUT 5799 Register 2132 slices | 55 MHz | 1/4.45 |
| | Image, Invisible/Visible, Pipe-lined | DCT | Xilinx Spartan 3 | 6810 LUT 6002 Register 2059 slices | 55 MHz | 1/4.45 |
| Ref [13] | Image, Invisible | Haar DWT | Xilinx VIRTEX-II | 3453 4inp LUT 1183 Flip flop 2145 slices | 29 MHz | 1/512 Bit per Pixel |
| | Video, Invisible | Simplified Haar DWT | Xilinx VIRTEX-II | 3153 LUT 190 Register Not reported | 3.91 ms/ frame | 1/512 Bit per Pixel |
| Proposed | Image, Invisible | Spatial | Spartan3 | 29 4inp LUT 16 slices | 13.7 ns | 1/9 Bit per Pixel |
| | Image, Invisible, Pipelined | Spatial | Xilinx VIRTEX-4 | 11019 4inp LUT 12191 Flip Flop 7888 slices | 172 MHz | 1/9 Bit per Pixel |